\documentclass[letter]{aa} 

\usepackage{graphicx}
\usepackage{txfonts}
\begin{document}

   \title{
   ALMA view of the circumstellar environment of the post-common-envelope-evolution binary system HD\,101584
   }

   \titlerunning{An ALMA view of HD\,101584}

   \author{H. Olofsson  \inst{1}
          \and
          W.H.T. Vlemmings \inst{1}
          \and 
          M. Maercker \inst{1}
          \and       
          E.M.L. Humphreys \inst{2}
          \and
          M. Lindqvist \inst{1}
          \and
          L. Nyman \inst{3,4}
          \and
          S. Ramstedt \inst{5}
          }

   \institute{Dept. of Earth and Space Sciences, Chalmers Univ. of Technology,
              Onsala Space Observatory, SE-43992 Onsala, Sweden\\
              \email{hans.olofsson@chalmers.se}
         \and
         ESO, Karl-Schwarzschild-Str. 2, D-85748 Garching bei M{\"u}nchen, Germany
         \and
         Joint ALMA Observatory, Alonso de Cordova 3107, Vitacura, Santiago de Chile, Chile
         \and
         ESO, Alonso de Cordova 3107, Vitacura, Santiago, Chile 
         \and
         Dept. of Physics and Astronomy, Uppsala University, Box 516, SE-75120 Uppsala, Sweden     
             }

   \date{Received 5 March 2015; accepted 24 March 2015}

 \abstract{}{}{}{}{} 
 
  \abstract
   {}
   {We study the circumstellar evolution of the binary HD101584, consisting of a post-AGB star and a low-mass companion, which is most likely a post-common-envelope-evolution system.}
   {We used ALMA observations of the $^{12}$CO, $^{13}$CO, and C$^{18}$O $J$\,=\,\mbox{2--1} lines and the 1.3\,mm continuum to determine the morphology, kinematics, masses, and energetics of the circumstellar environment. }
   {The circumstellar medium has a bipolar hour-glass structure, seen almost pole-on, formed by an energetic jet, $\approx$\,150\,km\,s$^{-1}$. We conjecture that the circumstellar morphology is related to an event that took place $\approx$\,500\,yr ago, possibly a capture event where the companion spiraled in towards the AGB star. However, the kinetic energy of the accelerated gas exceeds the released orbital energy, and, taking into account the expected energy transfer efficiency of the process, the observed phenomenon does not match current common-envelope scenarios. This suggests that another process must augment, or even dominate, the ejection process. A significant amount of material resides in an unresolved region, presumably in the equatorial plane of the binary system.}
   {}

   \keywords{Circumstellar matter --
          Stars: individual: HD101584 --
          Stars: mass-loss --
          Stars: AGB and post-AGB -- 
          Radio lines: stars
               }

   \maketitle
%

\section{Introduction}

\object{The source HD\,101584} (V885~Cen, IRAS\,11385--5517) is bright at optical
wavelengths ($V$\,$\approx$\,7 mag) with a spectral type A6Ia \citep{sivaetal99,kipp05}.  
\citet{bakketal96a} presented optical and infrared
data that indicate that it evolved from the asymptotic giant branch (AGB) at most a few 100 years ago, and they estimated a (present-day) mass of $\approx$\,0.6\,$M_\odot$, a luminosity of $\approx$\,4000\,$L_\odot$, and a distance of $\approx$\,0.7\,kpc.

Photometric and radial velocity variations show that HD\,101584 has a
binary companion. \citet{bakketal96b} and \citet{diazetal07} found periods of 218 and 144 days, respectively. \citet{bakketal96b} inferred an essentially edge-on circumbinary disk to explain optical absorption lines. The
absence of spectroscopic emission from the companion indicates that this is a low-mass 
main-sequence star. 

Images from the Hubble Space Telescope (HST) show a diffuse circumstellar environment. There may be a ring of radius $\approx$\,1\farcs5 roughly centred on the star \citep{sahaetal07}, that is almost
circular, which suggests that we see it almost pole-on. The central star
is bright, presumably because material around the polar axis of the system has partially been evacuated.

The circumstellar gas has been detected in molecular line emission.  \citet{olofnyma99} obtained
$^{12}$CO and $^{13}$CO $J$\,=\,\mbox{1--0} and \mbox{2--1} single-dish map data. 
They found at least three
kinematic components: a low-velocity, an intermediate-velocity, and a high-velocity
component of total widths $\approx$\,15, $\approx$\,100, and $\approx$\,300\,km\,s$^{-1}$, respectively.
The former two were spatially unresolved, and the high-velocity gas shows
an east-west bipolar outflow with a Hubble-like velocity gradient, the most blue- and red-shifted emission 
at $\approx$\,5$\arcsec$ to the W and the E.

A narrower ($\approx$\,80\,km\,s$^{-1}$), double-peaked OH 1667\,MHz maser line was mapped by \citet{telietal92}. The integrated OH emission is centred on the star (within 0\farcs3),
but is absent inside a radius of $\approx$\,1$\arcsec$. The
velocity increases systematically along the position angle
(PA) $\approx$\,--60$^\circ$ with the most blue- and red-shifted emission at $\approx$\,2$\arcsec$
to the SE and the
NW. OH emission and a 10$\,\mu$m feature \citep{bakketal96b} indicate an O-rich (C/O$<$1)
circumstellar medium.

Hence, the circumstellar environment of HD\,101584 appears to
be consistent with that of a classic proto-planetary nebula (proto-PN): an equatorial density enhancement with
an orthogonal, bipolar, high-velocity outflow \citep[e.g.,][]{alcoetal01, alcoetal07, castetal02, sahaetal06}. 

In this {\it Letter} we present data acquired with the Atacama Large Millimeter/sub-millimeter Array\footnote{This Letter makes use of the following ALMA data: ADS/JAO.ALMA\#2012.1.00248.S. ALMA is a partnership of ESO (representing its member states), NSF (USA) and NINS (Japan), together with NRC (Canada) and NSC and ASIAA (Taiwan), in cooperation with the Republic of Chile. The Joint ALMA Observatory is operated by ESO, AUI/NRAO and NAOJ.} (ALMA) for the $^{12}$CO, $^{13}$CO, and C$^{18}$O $J$\,=\,\mbox{2--1} line  and the 1.3\,mm continuum  on HD101584. In addition to this, we have detections of line emission from SiO,  CS, H$_2$S, SO, SO$_2$, OCS, H$_2$CO, and some of their isotopologues, which will be presented in forthcoming publications.


\begin{table*}
\caption{Observing parameters}
\centering
\begin{tabular}{l c c c c c c c c}
    \hline \hline
    Tuning        & $\nu$ & $t_{\rm on-source}$ & $\Delta V$ & channel rms & Beam   & Flux cal. & J1107-4449 & J1131-5818 \\
                  &  [GHz] & [min] & [km\,s$^{-1}$] & [mJy\,beam$^{-1}$] & $[\arcsec\times\arcsec, ^\circ]$  & & Flux [mJy] & Flux [mJy] \\
   \hline
$^{12}$CO & 230.53 & 8 & 0.64 & 1.8 & $0.65\times0.57, 8.2$   & Ceres & $710\pm0.7$ & $92.4\pm0.3$ \\
$^{13}$CO,  C$^{18}$O& 220.06 & 26 & 0.67 & 0.8 & $0.65\times0.54, 46.1$   & Titan & $680\pm2$ & $86.3\pm0.3$\\
  \hline
\end{tabular}
\label{t:observations}
\end{table*}

\section{Observations}

The $^{12}$CO, $^{13}$CO, and C$^{18}$O $J$\,=\,\mbox{2--1} line data were obtained on 29 April 2014 with 35 antennas of the ALMA 12\,m array in two frequency settings in band 6 (cycle\,1). In both cases, the data contain four spectral windows (spw) of 1.875\,GHz width and with 3840 channels. The observing parameters of the spw relevant for the data presented here are listed in Table~\ref{t:observations}.

The baselines range from 16 to 497\,m. This means a maximum recoverable scale of $\approx$\,10\arcsec . Bandpass and gain calibration were performed on J1107-4449 and J1131-5818 ($3^\circ$ separation from the target). Flux calibration was made using Ceres and Titan for the different tunings (Butler-JPL-Horizons 2012, ALMA Memo 594). Based on the calibrator fluxes determined in the two tunings, we estimate the absolute flux calibration to be accurate to within 5\%.

The data were reduced using CASA 4.2.2. After corrections for the time and frequency dependence of the system temperatures and rapid atmospheric variations at each antenna using WVR data, bandpass and gain calibration were made. Poor phase coherence led to the flagging of three antennas in the $^{12}$CO tuning and one antenna in the $^{13}$CO tuning. Subsequently, for each individual tuning, self-calibration was performed on the strong continuum. Imaging was made using the CASA clean algorithm after a continuum subtraction was performed on the emission line data. Our final line images were created using natural weighting and the restoring beam given in Table~\ref{t:observations}. We produced a single continuum map combining both tunings with a line-free equivalent bandwidth of 10.2\,GHz.


\section{Observational results and discussion}

\subsection{Global characteristics}

\begin{figure}
   \centering
   \includegraphics[width=6cm]{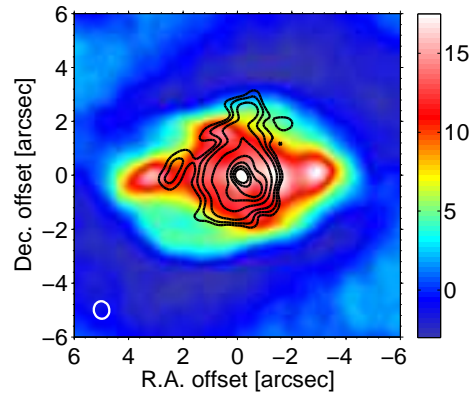}
      \caption{1.3\,mm continuum image (black contours) of HD\,101584 superposed on the global $^{12}$CO(2--1) map (synthesized beam in the lower left corner). The dynamic-range-limited rms is 0.14\,mJy\,beam$^{-1}$ in a beam of 0\farcs56$\times$0\farcs49 (PA\,=\,41.4$^\circ$) in the continuum. The contour levels are (2, 4, 8, 20, 40, 60, 80, 100)\,$\times$\,rms. The line intensity scale unit is Jy\,km\,s$^{-1}$\,beam$^{-1}$.
              }
         \label{f:cont}
   \end{figure}

The 1.3\,mm continuum map, overlaid on the global $^{12}$CO(2--1) map, is shown in Fig.~\ref{f:cont}. The total flux density, $S_{230}$\,=\,112\,mJy, falls short by almost a factor of four compared to the extrapolated spectral energy distribution (SED) model of \citet{bakketal96a}, suggesting a significant loss of flux originating from a region larger than 10\arcsec . The morphology is complex with faint structures surrounding an intense inner region, where $\approx$\,70\% of the total flux comes from inside a central 2\arcsec\, diameter circle.


\begin{figure}
   \centering
   \includegraphics[width=6cm]{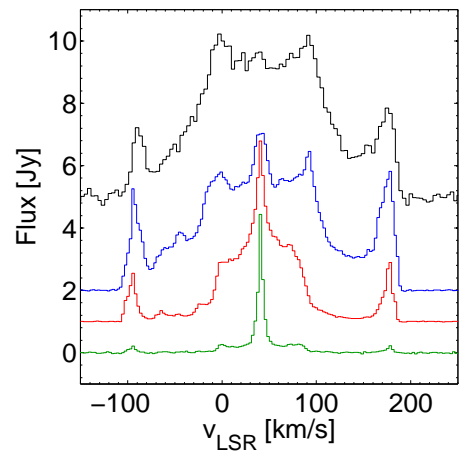}
      \caption{SEST $^{12}$CO (black, offset +5\,Jy) and the ALMA global $^{12}$CO (blue, offset +2\,Jy), $^{13}$CO (red, offset +1\,Jy, scaled $\times$1.5), and C$^{18}$O (green, scaled $\times$3) $J$\,=\,\mbox{2--1} lines.
              }
         \label{f:12co_globalprofile}
   \end{figure}

The ALMA global CO line profiles are shown in Fig.~\ref{f:12co_globalprofile}. The systemic velocity, $\upsilon_{\rm sys}$, is estimated to be 42\,$\pm$\,1\,km\,s$^{-1}$ (LSR). The ALMA total fluxes, integrated over the global line profiles, are 690, 150, and 18\,Jy\,km\,s$^{-1}$ for the $^{12}$CO, $^{13}$CO, and C$^{18}$O $J$\,=\,\mbox{2--1} lines, respectively. The isotopologue line intensity ratios in the full velocity range and the central 20\,km\,s$^{-1}$ are $^{12}$CO/$^{13}$CO\,$\approx$\,5 and 2 (solar value 89) and $^{13}$CO/C$^{18}$O\,$\approx$\,8 and 4 (solar value 6), respectively. This indicates that opacity effects are significant in the $^{12}$CO line, affect the $^{13}$CO line, and that they increase significantly towards the central velocity range. Assuming that the $^{16}$O/$^{18}$O ratio is solar (reasonable for a lower-mass star), the $^{13}$C may not be too far from solar.

The single-dish $^{12}$CO(\mbox{2--1}) flux is 850\,Jy\,km\,s$^{-1}$ \citep{olofnyma99}, with an uncertainty of $\approx$\,20\%, that
is, some flux is probably missed by ALMA. Most likely, based on a detailed comparison of the ALMA global and the single-dish line profiles, no flux is missed in the central 20\,km\,s$^{-1}$ and at the extreme velocity peaks, while $\approx$\,25\% of the flux is missed in the velocity range 10\,<\,$\mid\upsilon - \upsilon_{\rm sys}\mid$\,<\,90\,km\,s$^{-1}$, Fig.~\ref{f:12co_globalprofile}. This suggests the presence of extended gas that has been efficiently accelerated.

\begin{figure}
   \centering
   \includegraphics[width=8.5cm]{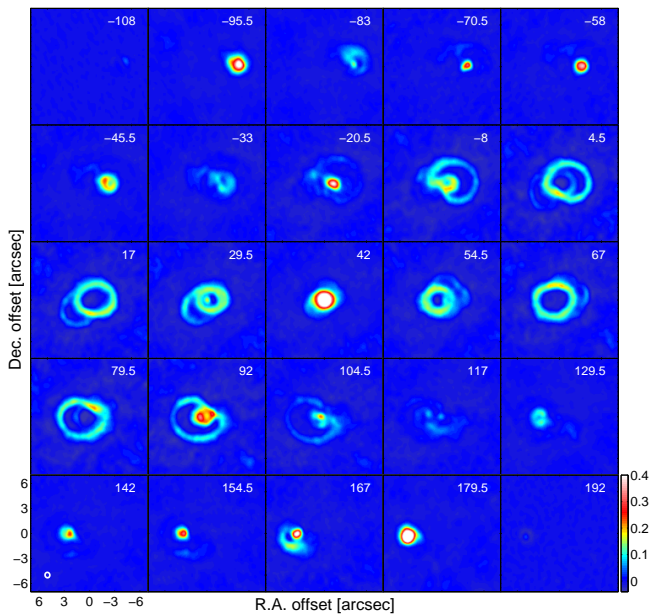}
      \caption{Velocity-channel maps ($\Delta \upsilon$\,=\,12.5\,km\,s$^{-1}$) of the $^{12}$CO($J$\,=\,\mbox{2--1}) emission towards HD\,101584 (synthesized beam in the lower left corner). The intensity scale unit is Jy\,beam$^{-1}$ (noise rms 2\,mJy\,beam$^{-1}$).
              }
         \label{f:12co_largemap}
   \end{figure}

The $^{12}$CO channel maps confirm the conclusion by \citet{olofnyma99} of an E-W oriented high-velocity outflow, but also show additional considerable complexity, Fig.~\ref{f:12co_largemap}. The morphology of the emission is very symmetric with respect to the systemic velocity. The high-velocity outflow is narrow and directed at PA\,$\approx$\,90$^\circ$. A complex ring-like structure is present in the velocity range --30\,--\,120\,km\,s$^{-1}$. The strong central emission is confined to a narrow velocity range, $\le\,20$\,km\,s$^{-1}$.

\subsection{A jet, an hour-glass structure, and the central region}


A number of different structures can be identified in different velocity ranges.

\begin{figure}
   \centering
   \includegraphics[width=6cm]{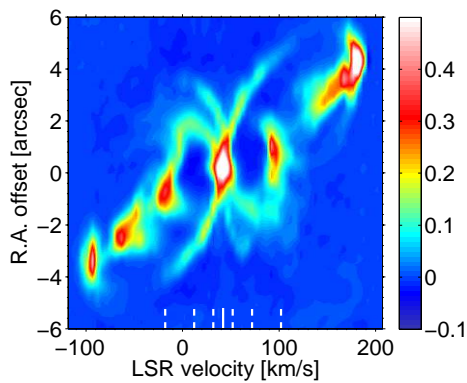}
      \caption{$^{12}$CO($J$\,=\,\mbox{2--1}) position-velocity diagram along PA\,=\,90$^\circ$. The systemic velocity (solid) and the velocity ranges discussed in text (hashed) are indicated. The intensity scale unit is Jy\,beam$^{-1}$.
              }
         \label{f:12co_posvel}
   \end{figure}

{\it $\mid$\,$\upsilon$-$\upsilon_{\rm sys}$\,$\mid$\,>\,60\,km\,s$^{-1}$}: The $^{12}$CO position-velocity (PV) diagram along the PA\,=\,90$^\circ$ shows a jet-like structure, with a Hubble-like velocity gradient, reaching a peak velocity of $\approx$\,150\,km\,s$^{-1}$, Fig.~\ref{f:12co_posvel}. The emission is patchy along the jet, which ends in two sharp features at the extreme velocities. The PNe BD+30$^\circ$3639 provides another example of such features \citep{bachetal00}. These may indicate that this is as far as the jet has travelled since its formation, or it has been stopped by surrounding material, or, at these distances, the jet is directed along the line of sight as a result of precessing, for example. 

\begin{figure}
   \centering
   \includegraphics[width=8.5cm]{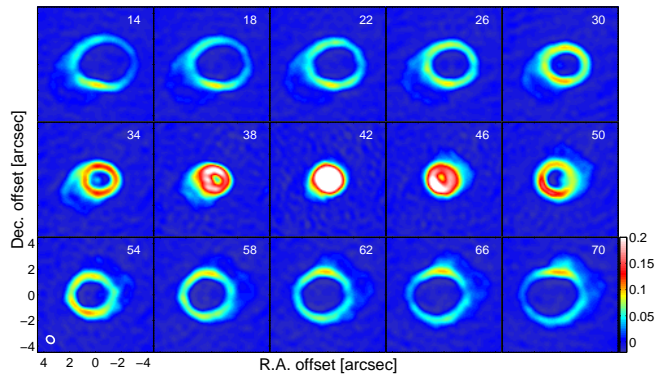}
      \caption{Velocity-channel maps ($\Delta \upsilon$\,=\,4\,km\,s$^{-1}$) of the $^{13}$CO($J$\,=\,\mbox{2--1}) emission towards HD\,101584 (synthesized beam in the lower left corner). The intensity scale unit is Jy\,beam$^{-1}$ (noise rms 2\,mJy\,beam$^{-1}$).
              }
         \label{f:13co_smallmap}
   \end{figure}

{\it $\mid$\,$\upsilon$-$\upsilon_{\rm sys}$\,$\mid$\,<\,30\,km\,s$^{-1}$}: The CO emission in the velocity range $\upsilon_{\rm sys}$\,$\pm$\,30\,km\,s$^{-1}$ is dominantly ring-like, as exemplified by the $^{13}$CO data in Fig.~\ref{f:13co_smallmap}. The ring size increases with offset from the systemic velocity, and its centre position depends on velocity in a manner consistent with the direction of the velocity gradient of the high-velocity outflow. The $^{12}$CO PV diagram, Fig.~\ref{f:12co_posvel}, shows the expected morphology of a hollow structure, slightly tilted w.r.t. the line of sight at PA\,$\approx$\,90$^\circ$ in this velocity range, suggesting an hour-glass structure seen almost pole-on. We estimate the inclination angle, $i$, to be about half of the opening angle of the hour-glass structure, $\approx$\,10$^\circ$. Surprisingly, the ring structure is elongated along PA\,$\approx$\,90$^\circ$, as opposed to the expected PA\,$\approx$\,0$^\circ$. 

At lower brightness level there is evidence for another bipolar structure in this velocity range. It has a velocity gradient opposite to that of the major CO outflow and a different position angle (PA\,$\approx$\,--45$^\circ$), Figs.~\ref{f:12co_largemap} and \ref{f:12co_posvel}. The direction of the structure and its velocity gradient are roughly consistent with those of the OH 1667\,MHz line.

{\it 30\,$\le$\,$\mid$\,$\upsilon$-$\upsilon_{\rm sys}$\,$\mid$\,$\le$\,60\,km\,s$^{-1}$}: At $\approx$\,$\pm$30\,km\,s$^{-1}$ from $\upsilon_{\rm sys}$, the ring-like emission breaks up into a complex structure, yet very symmetric w.r.t. the centre, Fig.~\ref{f:12co_largemap}. The structure ends at offset velocities of $\approx$\,60\,km\,s$^{-1}$ in two compact features displaced by $\approx$\,1\arcsec\,  on either side of the centre, along PA\,$\approx$\,90$^\circ$. Assuming that this structure has its origin in a region close to the binary system, and adopting a tilt angle of 10$^\circ$, a projected expansion length of 1${\farcs}$5, and an expansion velocity of 60\,km\,s$^{-1}$, the age is estimated to be $\approx$\,500\,yr. With the same assumption on the tilt angle, the time scale of the jet is $\approx$\,520\,yr. Thus, the complex structure and the jet could have their origins in the same event. 

{\it $\mid$\,$\upsilon$-$\upsilon_{\rm sys}$\,$\mid$\,<\,10\,km\,s$^{-1}$}: Using the C$^{18}$O({2--1}) line as a mass indicator, it is clear that a substantial fraction of the mass lies in the central region. About 
55\% of the integrated flux comes from the range $\upsilon_{\rm sys}$\,$\pm$\,10\,km\,s$^{-1}$. Further evidence for substantial mass in this region is the strong and narrow  H$_2$S(\mbox{2$_{20}$--2$_{11}$}) line, and the corresponding lines from its $^{33}$S and $^{34}$S isotopologues. Part of the emission in this velocity range comes from a circular region $\approx$\,2\arcsec\, in diameter (possibly related to the ring in the HST images), and part of the emission comes from an unresolved component at the center. The unresolved component is characterized by narrow spectral features, only a few km\,s$^{-1}$ wide. It is prominent in high-energy ($E_{\rm up}$\,$\approx$\,100\,K) SO$_2$ lines, for instance.

\subsection{Mass, energy, and momentum estimates}

A simple estimate of the dust mass using
\begin{equation}
\label{e:dustmass}
M_{\rm d} = \frac{S_{230} D^2}{\kappa_{230} B_{230}}\,\, ,
\end{equation}
where $B_{230}$ is the blackbody brightness at 230\,GHz, with $S_{230}$\,=\,112\,mJy, $D$\,=\,0.7\,kpc, the dust opacity $\kappa_{230}$\,=\,0.5\,cm$^2$\,g$^{-1}$ \citep{liseetal15}, and an adopted dust temperature of 45\,K [from the SED in \citet{bakketal96a}] gives $M_{\rm d}$\,$\approx$\,0.007\,$M_\odot$ ({with an uncertainty of, at least, a factor of four). This suggests about a solar mass of gas in the central region, assuming a dust-to-gas ratio of 200. There probably is considerably more mass in an extended region [\citet{bakketal96a} estimated 0.02\,$M_\odot$ in total].

A simple estimate of the gas mass is obtained using
\begin{equation}
\label{e:gasmass}
M_{\rm g} = \frac{16\pi m_{\rm H}}{hc g_{\rm u} A_{\rm ul} f}\,
     I\, D^2\, Q(T_{\rm rot})\, e^{E_{\rm u}/kT_{\rm rot}}
,\end{equation}
where normal symbols are used for the constants, and $f$
is the abundance with respect to H$_2$, $I$ the flux density integrated over velocity range, $Q$ the 
partition function, $T_{\rm rot}$ the rotational temperature, and 
$E_{\rm u}$ the upper level energy.  Using the 
C$^{18}$O(\mbox{2--1}) line, $I$\,=\,18\,Jy\,km\,s$^{-1}$, $f_{\rm C^{18}O}$\,=4\,$\times$10$^{-7}$
[solar $^{16}$O/$^{18}$O, $f_{\rm CO}$\,=\,2$\times$10$^{-4}$ (O-rich envelope)], and $T_{\rm rot}$=30\,K \citep{bujaetal01}, we obtain $M_{\rm g}$\,$\approx$\,0.3\,$M_\odot$ (with some considerable uncertainty, e.g., the gas temperature probably varies over the source). More than half of it lies in the central 20\,km\,s$^{-1}$ range.

Using the C$^{18}$O(\mbox{2--1}) line and the same assumptions as above, we find that the kinetic energy in the material accelerated to velocities $\ge$\,10\,km\,s$^{-1}$ is $\approx$\,4\,$\times$\,10$^{45}$\,erg (a lower limit since extended flux from accelerated gas is missed). 
This lies in the middle of the range found by \citet{bujaetal01}, 10$^{44-47}$\,erg, for the bipolar outflows of a sample of 28 proto-PNe.

Likewise, we find a momentum of $\approx$\,10$^{39}$\,g\,cm\,s$^{-1}$ in the accelerated gas, which lies in the middle of the range found by \citet{bujaetal01}, 10$^{37-40}$\,g\,cm\,s$^{-1}$. The estimated momentum rate over 500\,yr is $\approx$\,7\,$\times$\,10$^{28}$\,erg\,cm$^{-1}$, which is substantially higher than that available from radiation ($L/c$), $\approx$\,5\,$\times$\,10$^{26}$\,erg\,cm$^{-1}$, a common situation for proto-PNe \citep{bujaetal01}. Hence, a different momentum source is required.

\subsection{A scenario}

Adopting the orbital parameters (period $P$\,=\,218 days and velocity semi-amplitude $K_1$\,=\,3\,km\,s$^{-1}$) and the mass estimate, $M_1$\,=\,0.6\,$M_\odot$, of the post-AGB star \citep{bakketal96b, bakketal96a}, $i$\,=\,10$^\circ$, and a circular orbit (reasonable if we have a capture event), we derive a mass function of 
\begin{equation}
\frac{(M_2 \sin i)^3}{(M_1 + M_2)^2} = \frac{4\pi^2 (a_1 \sin i)^3}{G P^2} = 6\times10^{-4}\,M_\odot
\end{equation}
for the binary system, where $M_2$ is the mass of the companion ($a_1$\,=\,$K_1$$P$/2$\pi$$\sin i$). Thus, we estimate that $M_2$\,$\approx$\,0.6\,$M_\odot$ at a distance of $a$\,$\approx$\,0.7\,AU ($a_2$\,=\,$a_1$$M_2$/$M_1$, $a$\,=\,$a_1$+$a_2$) to the primary. This could be the result of a capture event if the companion started at a distance sufficiently nearby for the AGB star to engulf it. The amount of orbital energy released is 
\begin{equation}
E_{\rm rel} = -\frac{G\,M_{1,i}\,M_2}{2a_{\rm i}} + \frac{G\,M_{1}\,M_2}{2a} \approx\,2\times10^{45}\,{\rm erg}
,\end{equation}
where $M_{\rm 1,i}$\,=\,1.6\,$M_\odot$ (assuming that the present circumstellar material of $\approx$\,1\,$M_\odot$ comes from the AGB star), and assuming $a_{\rm 2,i}$\,=\,4\,AU (sufficiently near to be captured). 

There are considerable uncertainties in the orbital and energy estimates, and the common-envelope evolution scenario is complex with an uncertain energy transfer efficiency \citep{ivanetal13}. But there is good evidence that we see the effect of a capture event in which a fair fraction of the mass of the AGB star was released as the companion spiraled inwards. Most of it remains in the equatorial plane of the binary, but some of it has been accelerated up to high velocities in the form of a bipolar jet. The complex structure at the end of the hour-glass structure may be the imprint of what took place during the capture. However, taken at face value, the released orbital energy is not enough, and another mechanism may augment, or even dominate, the ejection event, for instance a magnetic field of the (rapidly) spinning [$\upsilon \sin i$\,$\approx$\,50\,km\,s$^{-1}$, \citet{bakketal96b}] post-AGB star.


\section{Conclusions}

We have presented CO isotopologue emission line data on the post-AGB object HD\,101584 obtained with ALMA and provided an initial analysis of the morphology, kinematics, masses, and energetics of the circumstellar medium. We conclude that the circumstellar medium of HD\,101584 has been severely affected by the evolution of the binary system. A high-velocity jet ($\approx$\,150\,km\,s$^{-1}$), directed at PA\,$\approx$\,90$^\circ$ and slightly tilted w.r.t. the line of sight, has excavated an hour-glass structure in the circumstellar medium. A complex structure is seen at the extreme velocities of the hour-glass structure. The estimated ages of the complex structure and the jet are comparable, $\approx$\,500\,yr. A common origin of these phenomena such as the infall of the companion is therefore possible. However, taken at face value, the kinetic energy of the accelerated gas exceeds the released orbital energy during the capture event, and, further taking into account the expected efficiency of the process, the observed phenomenon does not match current common-envelope scenarios. This suggests that another process must augment, or even dominate, the ejection process. There appears to be substantial amount of material, maybe as much as a solar mass, in the central region, which is characterized by narrow spectral features that are only a few km\,s$^{-1}$ wide. This may be the gas removed from the AGB star when the companion spiraled inwards. At low brightness levels there are indications of another bipolar structure. Its velocity gradient is opposite to that of the CO outflow and directed at PA\,$\approx$\,--45$^\circ$. 

\begin{acknowledgements}
We are deeply grateful to the late Patrick Huggins, who played an instrumental role in defining this project. HO and WV acknowledge support from the Swedish Research Council. WV acknowledges support from the ERC through consolidator grant 614264. MM has received funding from the People Programme (Marie Curie Actions) of the 
EU's FP7 (FP7/2007-2013) under REA grant agreement No. 623898.11.
\end{acknowledgements}


\end{document}